\documentclass[a4paper,11pt]{article}
\usepackage{jheppub}

\usepackage{graphicx}
\usepackage{bm}
\usepackage{hyperref}%
\usepackage{amssymb}
\usepackage{amsmath}
\usepackage{amstext}
\usepackage{amsfonts}
\usepackage{amssymb}
\usepackage{subfigure}
\usepackage{color}
\usepackage{epsfig}
\usepackage{bbm}

\newcommand{\fm}{\mbox{fm}}

\newcommand{\Mev}{\mbox{MeV}}
\newcommand{\Gev}{\mbox{GeV}}

\newcommand{\maxv}{\mbox{max}}
\newcommand{\ov}{\mbox{ov}}
\newcommand{\plaq}{\mbox{pl}}
\newcommand{\rt}{\mbox{rt}}
\newcommand{\Tr}{\mbox{Tr}}
\newcommand{\sign}{\mbox{sign}}
\newcommand{\spec}{\mbox{spec}}

\title{Determination of the properties of vector
mesons in external magnetic field by Quenched $SU(3)$ Lattice QCD}

\author[a,b]{E.V. Luschevskaya}
\author[a]{O.E. Solovjeva}
\author[c,d]{O.V. Teryaev}
 
\affiliation[a]{Institute for Theoretical and Experimental Physics named by A.I.Alikhanov of NRC «Kurchatov Institute, 117218,  Bolshaya Cheremushkinskaya  25, Moscow, Russia}
\affiliation[b]{Moscow Institute of Physics and Technology, Dolgoprudnyj, Institutskij lane 9, Moscow Region 141700, Russia}
\affiliation[c]{Joint Institute for Nuclear Research, Dubna, 141980, Russia}
\affiliation[d]{Lomonosov Moscow State University, GSP-1, Leninskie Gory, 119991 Moscow, Russia}

\emailAdd{luschevskaya@itep.ru}
\emailAdd{olga.solovjeva@itep.ru}
\emailAdd{teryaev@theor.jinr.ru}
\abstract{
We investigate   the  ground state energies of   vector $\rho^{\pm}$ and $K^{\pm *}$ mesons depending on  the magnetic field value  in the $SU(3)$ lattice gauge theory. It has been shown that the energy of a vector particle   depends on its spin projection on the field axis. The magnetic dipole polarizability and hyperpolarizabilities give  significant contributions  to the energy value which prevents   the formation   of the charged vector meson condensate at high magnetic fields.  We calculate the g-factor of  $\rho^{\pm}$ and $K^{\pm*}$ mesons and the  dipole magnetic polarizability of $\rho^{\pm}$ mesons.  
}

\begin{document} 
\maketitle
\flushbottom


\section{Introduction}
\label{intro}

Magnetic
fields of hadronic scale could exist in cosmic objects  and Early Universe. The noncentral heavy ion collisions may create such fields in terrestrial laboratories like LHC (ALICE), RHIC, NICA and
FAIR (CBM) \cite{Skokov:2009}. The sufficiently strong   magnetic field affects the internal structure of hadrons, in particular shifts their energy levels. 

Since a meson is a composite particle, the external electromagnetic field will cause the deformation of the meson wave function. The way the meson is deformed under the action of the external field is determined by the properties of the strong interaction between the quark and the antiquark.

In this work we consider the constant external  magnetic field.
The
magnetic moments, the dipole magnetic polarizability and
the hyperpolarizabilities   characterize  the response of  a particle to
this external influence.  The magnetic moment is the most important quantity describing the magnetic properties of a meson or baryon due to the presence of the spin. The magnetic polarizabilities and  hyperpolarizabilities show  the
distribution of the quark currents inside an hadron  and describe its internal structure in the external magnetic field. 
 
The effects of a magnetic
field on an hadron mass  were firstly studied in the pioneering paper \cite{Martinelli:1982}. Then the magnetic field was introduced in two flavor lattice QCD simulation \cite{Delia:2010}  and  used as the probe of QCD properties \cite{Massimo:2015}.
 The energies of hadrons in magnetic fields have been calculated on the lattice \cite{Bali:2015,Luschevskaya:2015a,Luschevskaya:2015b,Savage:2015},  in  theoretical models \cite{Simonov:2013,Liu:2015,Taya:2015,Kawaguchi:2016,Hattori:2016} and  in the framework of the QCD sum rules \cite{Cho:2015,Gubler:2015}.
The notion of the  hadron polarizability was initially discussed  in 
 \cite{klein,baldin}. The  magnetic polarizabilities of pions have been measured in the experiments  
\cite{Antipov:1983,Filkov:2006,Adolph:2015},
have been calculated in the chiral perturbation theory
\cite{Gasser:2005,Aleksejevs:2013} and in the  lattice gauge theory \citep{Luschevskaya:2016}.
  The magnetic polarizabilities of
baryons  have been obtained in full lattice QCD  \cite{Savage:2015}. The
magnetic moment  of the $\rho$ meson  has been explored in
\cite{Andersen,Samsonov,Braguta,Zanotti,Simonov,Djukanovic,Owen}.

 This work is devoted to exploration of the   energy levels of $\rho$ and $K^*$ mesons in the magnetic field, their magnetic polarizabilities and  magnetic moments.   
Our results were obtained in pure $SU(3)$ lattice gauge theory. The inclusion of dynamical quarks into a consideration will most likely not lead to a radical change in the magnetic properties of $\rho$ meson, although it becomes an unstable particle in full QCD.
  For example, in 2+1 full QCD  the g-factor of the  $\rho^{\pm}$ meson  is equal to 2.21(8) at the physical point \cite{Owen}. Quenched lattice theory predicts the value $g=2.25(0.34)$ in the chiral limit \cite{Andersen} and $g=2.20(0.15)$ at the lowest pion mass \cite{Zanotti}.

In Section \ref{sec-1} we describe the technical details of our calculations, the gauge field action,  the fermionic spectrum and correlation functions. In Section \ref{sec-2} the energy of the charged vector
$\rho$ meson has been studied for various spin projections   versus  the magnetic field value. Section
\ref{sec-3} is devoted to the discussion of the magnetic moments of the $\rho^{\pm}$ and $K^{*\pm}$ mesons.
 The   magnetic dipole polarizability and hyperpolarizability of the first order  of the $\rho^{\pm}$ mesons have been calculated in Section \ref{sec-4}.

\section{Details of calculations}
 \label{sec-1}

\subsection{Fermionic spectrum}
\label{spectrum}

To calculate  the eigenvalues and the eigenvectors of the Dirac operator we use
the Neuberger overlap operator  \cite{Neuberger:1997}.
It has the following form
\begin{equation}
D_{\ov}=\frac{\rho}{a} \left( 1+\frac{D_W}{\sqrt{D^{\dagger}_W D_W}}  \right)=\frac{\rho}{a} \left( 1+\gamma_5\sign(H) \right),
\label{overlap}
\end{equation}
where $D_W=M-\rho/a$ is the Wilson-Dirac operator with the negative
mass term $\rho/a$, $\rho=1.4$ is the parameter,  $H=\gamma_5 D_W $ is the hermitian Wilson-Dirac operator, $a$ is the lattice spacing in
physical units, $M$ is the Wilson hopping term with $r=1$.
 The key ingredient of the overlap operator is the $\sign(H)$ function
 \begin{equation}
\sign(H)=\frac{H}{ \sqrt{H^{\dagger} H}}.
\label{sign_function}
\end{equation}
Let's construct the massive overlap operator  
 \begin{equation}
 M_{\ov}=\left(1-\frac{am_q}{2 \rho}\right) D_{\ov}+m_q
 \end{equation}
for the   quark mass $m_q$.

For the numerical implementation of this operator the MinMax polynomial   approximation of the Sign function is used \cite{Giusti:2003}.
The spectrum  of the   Dirac operator $\spec(H) \in [\lambda_{min},\lambda_{max}]\in {\cal R}$.
The Sign function also can be expressed through the norm $||H||$
 \begin{equation}
\sign(H)=\sign\left( \frac{H}{||H||}  \right)=\sign(W).
\label{sign_function2}
\end{equation}
Since $||H||=\lambda_{max}$,  $\spec(W)\in
[\lambda_{min}/\lambda_{max};1]$. The function $1/||H||$ is approximated  by the Chebyshev polynomials  $T_k(z)$, $k=0,..,n$   on the interval $\sqrt{\epsilon}\leqslant \spec(H)
\leqslant 1$, where $\epsilon=\lambda^2_{min}/\lambda^2_{max}$. The  original matrix $H$ and the
 polynomial  
\begin{equation}
P_n(H^2)=\sum_{k=0}^{n}c_k T_k(z),\ \ z=\frac{2H^2-1-\epsilon}{1-\epsilon},
\end{equation}  
  has the same  set of eigenfunctions
$\psi_k$ \cite{Neff:2001}. 

The maximal relative error is 
\begin{equation}
\delta=\maxv|h(y)|,
\end{equation}
where 
\begin{equation}
h(y)=\frac{1/||H||-P_n(H^2)}{1/||H||},
\end{equation}
in our calculations  $\delta \sim 10^{-8}$.
The fermionic propagators are calculated using the  eigenfunctions and the eigenvalues  of the overlap Dirac operator. 
  This method    controls the computational errors with a high efficiency and preserves the chiral invariance at a zero quark mass on the lattice
\cite{Giusti:2003}.

\subsection{Gauge field action}

We generate ensembles of $200-350$ statistically independent
quenched  SU(3)   configurations of the gauge field using
  the tadpole improved   L\"uscher-Weisz action \cite{Luscher:1985}.
\begin{equation}
S=\beta_{imp} \sum_{\plaq} S_{\plaq}-\frac{\beta_{imp}}{20 u^2_0}\sum_{\rt}S_{\rt},
\label{action}
\end{equation}
where $S_{\plaq,\rt}=(1/2)\Tr(1-U_{\plaq,\rt})$ is the plaquette (denoted by $\plaq$) or 1$\times$2 rectangular loop term ($\rt$),
$u_0=(W_{1\times1})^{1/4}=\langle(1/2)\Tr U_{\plaq}\rangle^{1/4}$ is the input tadpole factor computed at zero temperature \cite{Bornyakov:2005}.
This action  suppresses  ultraviolet dislocations which  leads to  non-physical near-zero modes of the Wilson-Dirac operator and difficulties in choice of the $\rho$ parameter (see Subsection \ref{spectrum}).

\subsection{Abelian magnetic field on the lattice}
\label{Calc.corr.func.} 
  
We consider  charged particles in a  constant external magnetic field.
The Abelian magnetic  vector potential $A_{\mu}$   is chosen  in the symmetric gauge
\begin{equation}
 A^B_{\mu}(x)=\frac{B}{2} (x_1 \delta_{\mu,2}-x_2\delta_{\mu,1}).
\end{equation}
This insertion  does not affect the dynamics of the gluon as we consider the quenched approximation.
The resulting gauge field is presented as the sum
\begin{equation}
A_{\mu \, ij}= A_{\mu \, ij}^{gl} + A_{\mu}^{B} \delta_{ij}
\label{gaugefield}
\end{equation}
of the nonabelian $SU(3)$ gauge field of gluons $A_{\mu \, ij}^{gl}$ and the $U(1)$ field of the magnetic
photons $A_{\mu}^{B}; i,j=1,.., N^2_c-1, \mu=1,2,3,4$ are the color and the Lorentz indices  respectively.

In order to respect the gauge invariance of the theory on the lattice and satisfy the periodic boundary conditions in space one has to   impose the twisted
boundary conditions on fermionic fields \cite{Al-Hashimi:2009}. 
In our case they have the following form
\begin{equation}
\psi(x_1+aN_s,x_2,x_3)=\exp(-i\frac{q}{2}B a N_s x_2)\psi(x_1,x_2,x_3),
\label{twbc1}
\end{equation}
\begin{equation}
\psi(x_1,x_2+a N_s,x_3)=\exp(i\frac{q}{2}B aN_s x_1)\psi(x_1,x_2,x_3),
\label{twbc2}
\end{equation}
where $N_s$ is the
number of lattice sites in spatial directions,   $q$ is the quark charge. There is also an additional gauge dependence of the electromagnetic gauge potential and fermion  field, which is analogous to the $\theta$-vacuum in QCD \cite{Al-Hashimi:2009}.

The non-Abelian gauge $SU(N)$ theories in a box   with an external electromagnetic field and periodic boundary conditions were originally investigated in \cite{Hooft:1979}. 
  The Landau level problem  has   been studied on a torus in papers \cite{Zainuddin:1989,Chen:1996}.
  The  condition of quantizing   for the magnetic flux can   be also derived from consistency of the boundary conditions  \eqref{twbc1} and  \eqref{twbc2}, see  \cite{Al-Hashimi:2009}. Therefore the magnetic field value   is  determined by the relation
\begin{equation}
qB=\frac{2\pi n_B}{(aN_s)^2}, \ \ n_B \in \mathbb{Z},
\label{quantization}
\end{equation}
where $q=-1/3e$ is the charge of the $d$-quark, $N_s$ is the size of the lattice  in the spatial direction.
   We explore the sufficiently large magnetic fields corresponding to $n_B\sim 0\div 28$, where the saturation regime  ($n_B/(L^2)\gtrsim 0.5$) is not  yet achieved. The radius of the first Landau level  $l_H=1/\sqrt{eB}$ is also larger for the range of fields and lattice spacings presented here.   The lattice parameters and the values of the magnetic field used for the calculations are presented in Appendix \ref{app}. 

\subsection{Calculation of correlation functions}
\label{Calc.corr.func.}

The correlation functions of the vector fermionic currents for the charged mesons  in the coordinate space
\begin{equation}
\langle O_{\rho^+}(x) \bar{O}_{\rho^+}(y)  \rangle=
 -\Tr[\Gamma_1D_u^{-1}(x,y)\Gamma_2D_d^{-1}(y,x)],
 \label{observables}
\end{equation}
 where $O_{\rho^+}=\psi^{\dagger}(x)_d \Gamma_{1,2} \psi(x)_u$ is the $\rho^+$ meson interpolator,  $\Gamma_1,\Gamma_2=\gamma_{\mu}$ are the gamma matrices, $D_{u,d}^{-1}$ are the   propagators of $u$ and  $d$  quarks.  The interpolation operator for the   $\rho^-$ meson is defined similarly, $O_{\rho^-}=\psi^{\dagger}(x)_u \Gamma_{1,2}\psi(x)_d$.
  The lattice interpolation operator for the $\pi^0$ meson has the form $O_{\pi^0}=(\psi^{\dagger}(x)_u \gamma_5\psi(x)_u-\psi^{\dagger}(x)_d \gamma_5\psi(x)_d)/\sqrt{2}$. The corresponding correlation function reads
 \begin{equation}
   \langle O_{\pi^0}(x) \bar{O}_{\pi^0}(y)  \rangle=
 -\frac{1}{2}\Tr[\gamma_5D_u^{-1}(x,y)\gamma_5D_u^{-1}(y,x)]-\frac{1}{2}\Tr[\gamma_5D_d^{-1}(x,y)\gamma_5D_d^{-1}(y,x)],
 \label{observables2}
\end{equation}
 where only the connected part is included. The disconnected part  is zero because the  $u$ and $d$ quarks contributions cancel each other. 

The massive Dirac propagator in  an external
magnetic field is expressed as the sum over   the lowest $M$ Wilson-Dirac eigenmodes
\begin{equation}
D^{-1}(x,y)=\sum_{k<M}\frac{\psi_k(x) \psi^{\dagger}_k(y)}{i \lambda_k+m},
\label{lattice:propagator}
\end{equation}
where
$x=(\textbf{n}a, n_ta)$ and $y=(\textbf{n}^{\prime}a, n^{\prime}_t a)$ are the coordinates on the lattice.
 $\textbf{n},\textbf{n}^{\prime}\in \Lambda_3=\{(n_1,n_2,n_3)|n_i=0,1,...,N_s-1\}$ denote the numbers of lattice sites in the spatial directions,   $n_t,n^{\prime}_t$ are the numbers of lattice sites in the time direction.
We use $M=50$ to provide the high level of  convergence.

We carry out the Fourier transformation  of our meson interpolators to the momentum space over the spatial components $\textbf{n}$
in the spatial lattice $\Lambda_3$. Then the interpolators are projected to zero spatial momentum
   $\textbf{p} =0$ which corresponds to the meson at rest. 
At zero magnetic field   the meson energy  equals to its mass $E_0=m_0$.

The correlation function in the Euclidean space can be represented as the series over   Hamiltonian eigenstates \cite{Lang:2010}:
 $$
C(n_t)=\langle \psi^{\dagger}(\textbf{0},n_t) \Gamma_1 \psi(\textbf{0},n_t) \psi^{\dagger}(\textbf{0},0) \Gamma_2 \psi(\textbf{0},0)\rangle_A =
 $$
\begin{equation}
\sum_k\langle 0|\hat{O}|k \rangle \langle k|\hat{O}^{\dagger}|0 \rangle e^{-n_t a E_k},
\label{sum}
 \end{equation}
 where   $E_k$ is the energy of the state with the quantum number $k$, $\hat{O}, \hat{O}^{\dagger}$ are  the  operators acting on the  Hilbert space.

The correlator is the second rank tensor invariant w.r.t. Lorentz group transformation.
It is, in fact, proportional to the vector meson spin covariant density matrix, which is  
especially convenient to consider in its rest frame. The symmetry is than reduced to rotational one and the irreducible representations include the scalar (corresponding to spin-averaged cross-section), axial vector (corresponding to 3 components of vector polarization) and symmetric traceless tensor (corresponding to 5 components of tensor polarization). 
 The background magnetic field is directed along the z axis. 
 It is convenient to express the spin density matrix in terms of transverse ($e_x=(0,1,0,0), e_y=(0,0,1,0)$) and longitudinal ($e_z=(0,0,0,1)$) polarization vectors.
   One can obtain the energy of the ground state for the spin projection $s_z=0$ on the field direction from equation \eqref{observables}, when $\Gamma_1,\Gamma_2=3$.
 The combinations of the correlators
    \begin{equation}
C(s_z=\pm 1)= C_{11}+C_{22} \pm i(C_{12}-C_{21}).
\label{eq:CVV1}
    \end{equation}
    give the energies of vector mesons  with the   spin projections equals to $+1$ and $-1$ correspondingly, indices '$1$' and '$2$' correspond to the $\gamma_1$ and $\gamma_2$ in \eqref{observables} respectively.
    
At large $n_t$ the first term   in   formula \ref{sum}, containing   the energy of the ground state, gives the main contribution to the correlation function. Due to the   boundary conditions it has the following form \cite{Lang:2010}:
\begin{equation}
C_{fit}(n_t)=2A_0 e^{-N_T a E_0/2} \cosh ((n_t-\frac{N_T}{2}) a E_0)
 \label{sum33}
\end{equation}
  where  $A_0$ is a constant, $E_0$ is the energy of the ground state.
We find   the ground state energy as the fit parameter, fitting   lattice correlators by the function \eqref{sum33}. 
In order to minimize the errors and exclude the contribution of excited states  the data are fitted at $n_0 \leq n_t \leq N_T-n_0$, where $n_0$ is the parameter which is determined from a plot of the effective mass.
The   errors are found from the  fit of the correlation function by  $\chi^2$ method.

The effective mass can be found from the formula
   \begin{equation}
   m_{eff}(n_t+\frac{1}{2})=  \ln{ \frac{C(n_t)}{C(n_t+1)}},
   \label{effmass_lat}
   \end{equation}
 but it doesn't take into account the boundary conditions.
  To respects  the periodicity  one has to solve     numerically  the following  equation
 \begin{equation}
\frac{C(n_t)}{C(n_t+1)}=   \frac{\cosh(m_{eff}(n_t-N_T/2))}{\cosh(m_{eff}(n_t+1-N_T/2))}.
\label{effmass_bc}
 \end{equation}

\begin{figure}[htb]
\begin{center}
\includegraphics[width=5.8cm,angle=-90]{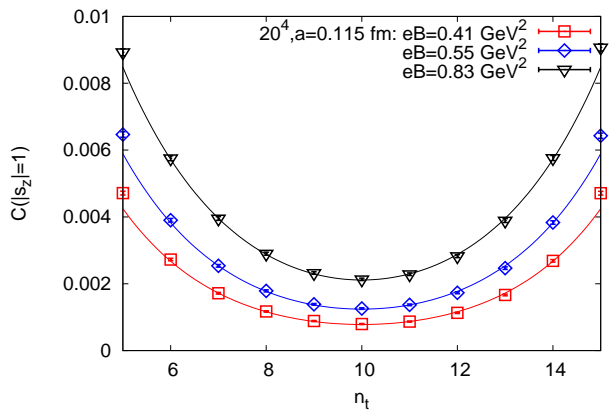}
\caption{The correlation function of  the   $\rho^{+}$ ($\rho^{-}$) meson   with the $s_z=+1$ ($s_z=-1$) spin projection  for   $m_{\pi}(B=0)=535(4)\ \Mev$ and several values of the magnetic field. The curves are the fits of the lattice data made by the hyperbolic cosine function  \eqref{sum33}.}
\label{corr}
\end{center}
\end{figure}

\begin{figure}[htb]
\begin{center}
\includegraphics[width=5.8cm,angle=-90]{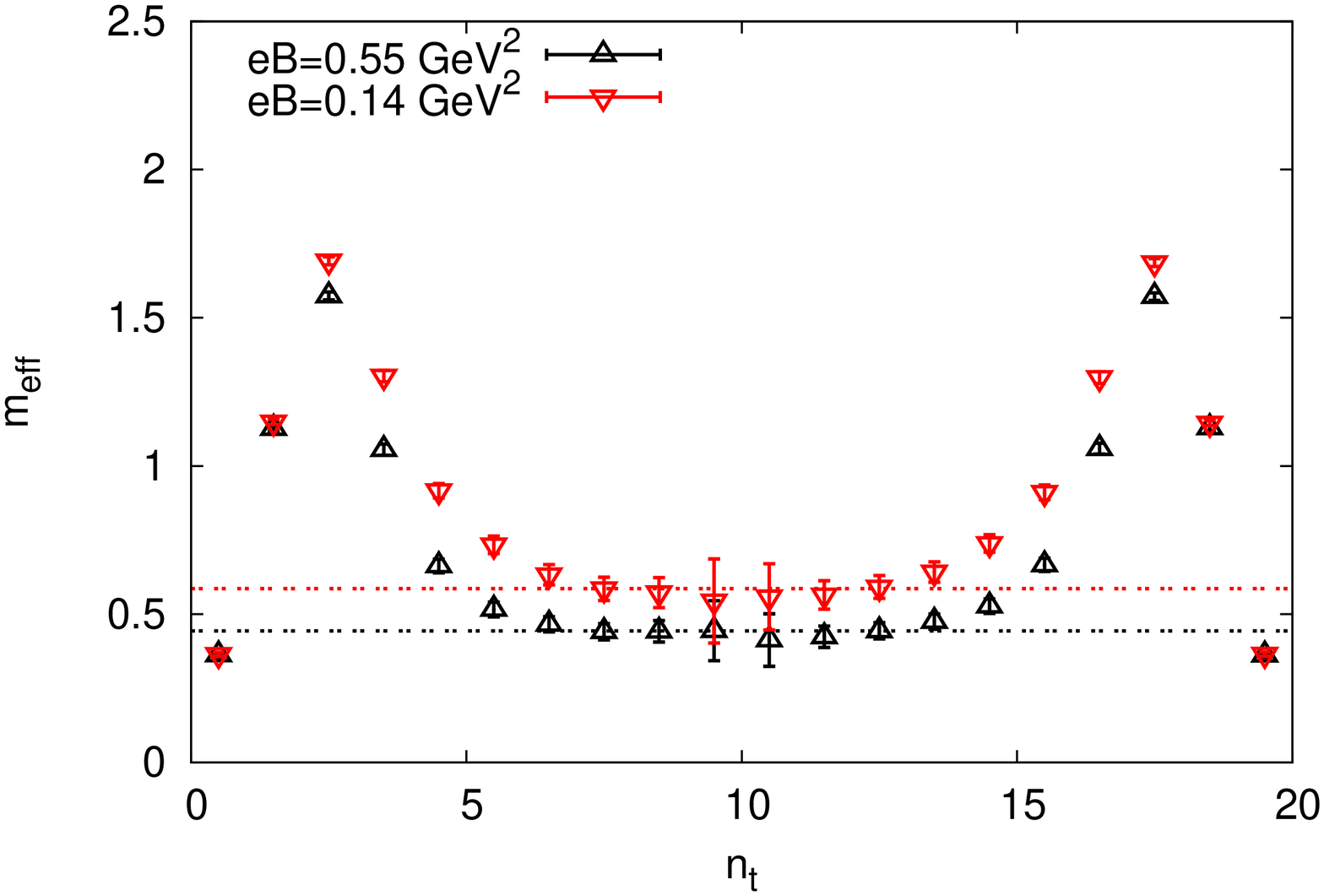}
\caption{The effective mass   of  the   $\rho^{+}$ ($\rho^{-}$) meson   with the $s_z=+1$ ($s_z=-1$) spin projection  for   $m_{\pi}(B=0)=535(4)\ \Mev$. The lines correspond to the  fits of the $m_{eff}$ by the constant function at $n_0=6$.}
\label{effmass_s-1}
\end{center}
\end{figure}

 Fig.\ref{corr}  shows the correlation function of  the $\rho^{+}$ ($\rho^{-}$) meson   with the $s_z=+1$ ($s_z=-1$) spin projection on the magnetic field direction for the lattice volume $20^4$, the lattice spacing $0.115\ \fm$, the pion mass $535(4)\ \Mev$ and several field  values. Fits were done using the  hyperbolic cosine function  \eqref{sum33} at $n_0=6$.
In Fig.\ref{effmass_s-1} the plot of the effective mass are shown   for the same lattice parameters  and for the magnetic field values $0.14\ \Gev^2$ and $0.55\ \Gev^2$.  The points represent the $m_{eff}$ values obtained from    equation \eqref{effmass_bc}, the lines correspond to the    fits by a constant function performed   at   $n_0=6$.  One can also use $n_0=5$, the energies obtained at $n_0=5$ and $n_0=6$ coincide within the errors.

\begin{figure}[htb]
\begin{center}
\includegraphics[width=5.8cm,angle=-90]{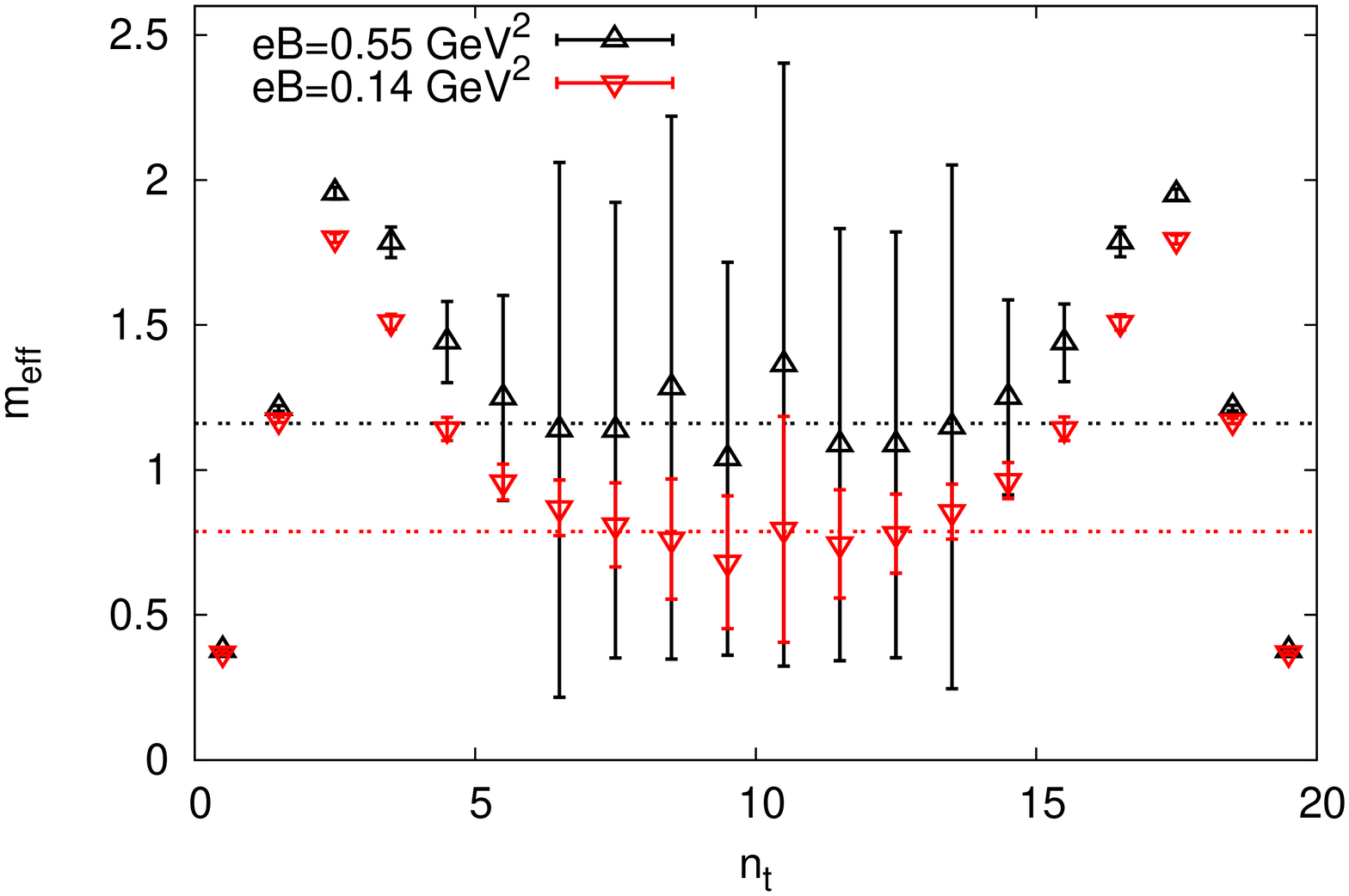}
\caption{The effective mass   of  the   $\rho^{+}$ ($\rho^{-}$) meson   with the $s_z=-1$ ($s_z=+1$) spin projection  for   $m_{\pi}(B=0)=535(4)\ \Mev$. The lines correspond to the  fits of the $m_{eff}$ by the constant function at $n_0=6$.}
\label{effmass_s+1}
\end{center}
\end{figure}

In Fig. \ref{effmass_s+1} we depict the effective mass plot  of the $\rho^{+}$ ($\rho^{-}$) meson for another  spin projection  $s_z=-1$ ($s_z=+1$) on the magnetic field direction.
 From a comparison of 
Fig. \ref{effmass_s-1} and Fig. \ref{effmass_s+1} we see that the errors of the mass determination are sufficiently larger for the highest energy sublevel than for the lowest sublevel.

\section{Magnetic moments of $\rho^{\pm}$ and $K^{*\pm}$ mesons}
\label{sec-2}

Let us estimate the Lande $g$-factor of the vector mesons from the  lattice data, using its values we obtain  the magnetic polarizabilities (Section \ref{sec-4}).
 The $g$-factor     characterizes  the  gyromagnetic ratio of a particle or its magnetic moment in natural magnetons.

  The vector $\rho^{\pm}$ and $K^{*\pm}$ mesons,  consisting of   the strongly interacting quarks and gluons, have a complex structure. 
The precise determination of the  $g$-factor  value   is of much interest because it enables to find the contribution of non-perturbative effects to the   magnetic moment of hadrons. However, the $\rho$ meson is an unstable and short-lived particle, so the experimental value of the $g$-factor is hard to measure.

The energy of the Landau levels  is described by the following formula
\begin{equation}
E^2=p^2_z+(2n+1)|qB|-gs_zqB+m^2, 
\label{eqLL}
\end{equation}
where $p_z$ is the momentum in the 'z' spatial direction, $n$ is the
principal quantum number, $q$ is the electric charge of the
meson, $g$ is the  g-factor, $s_z$ is the spin projection on the
field direction and $m\equiv E(B=0)$  is the energy of the particle
at zero magnetic field  and zero momentum. Further we consider $n=0$ and $p_z=0$.
This equation correspond to pint-like particle and should get the corrections from its structure for high mgnaetic fields. 

At relatively low magnetic fields ($ \leq 0.5$ $\Gev^2$)   the energy squared of the $\rho^{\pm}$ mesons iexhibits the linear responce of the field
\begin{equation}
E^2=|qB|-gs_zqB+m^2.
\label{eqLL1}
\end{equation}
One can neglect the    nonlinear response on the magnetic field because
the next term contribution to the energy at these $B$ values  is less then $10\%$ (see Section \ref{sec-3}). The energy of a meson can be described by a formula \eqref{eqLL1}, because  at the magnetic fields lower than QCD scale the internal structure of the meson has not to be revealed. It follows from equation $\eqref{eqLL1}$,  the energy  splits into sublevels in the external magnetic field. The $E^2$ value  increases at $qs_z=0,\,-1$ and decreases at $qs_z=+1$.  

\begin{figure}[htb]
\begin{center}
\includegraphics[width=5.8cm,angle=-90]{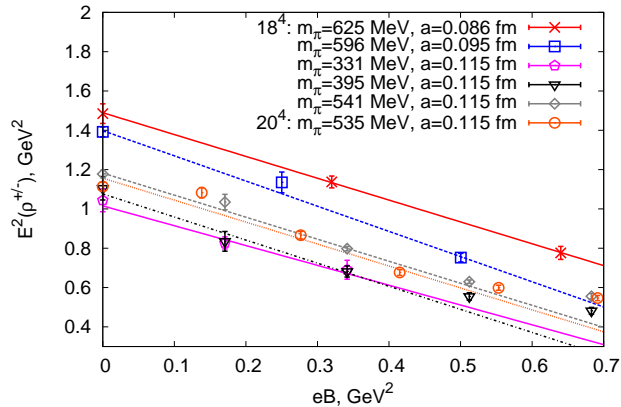}
\caption{The energy squared of the $\rho^{+}$ ($\rho^-$) meson with the spin projection $s_z=+ 1$ ($s_z=-1$) versus the magnetic field value for   various lattice volumes, lattice spacings  and different quark masses. The points correspond to the lattice data, the lines are the fits to these data obtained using formula \eqref{eqLL1}.}
\label{Fig:g:rho}
\end{center}
\end{figure}

In Fig.\ref{Fig:g:rho} we show the   energy squared of the $\rho^{+}$ ($\rho^{-}$) meson depending on the field value with the spin projection $s_z=+1$ ($s_z=-1$) for the lattice volume $18^4$,  the lattice spacings $0.086\ \fm$, $0.095\ \fm$, $0.115\ \fm$ and  various pion masses, the lowest pion mass is equal to $331(7)\ \Mev$ at $a=0.115\ \fm$.  
For the lattice volume $20^4$  the calculations were carried out at the pion mass $m_{\pi}=535(4)\ \Mev$ and the lattice spacing $a=0.115\ \fm$.
In Fig.\ref{Fig:g:rho} the points represent the lattice data, the lines are the fits to the lattice results obtained with the use of formula \eqref{eqLL1}. 

\begin{figure}[htb]
\begin{center}
\includegraphics[width=5.8cm,angle=-90]{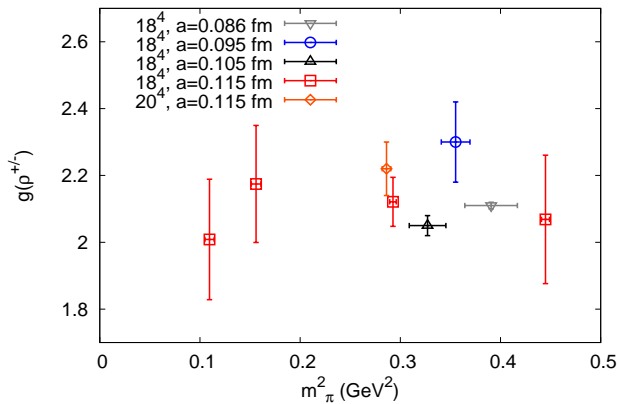}
\caption{The g-factor of the charged $\rho$ meson versus the mass squared of the  pion for various lattice data sets.}
\label{Fig:g:rho:mpi}
\end{center}
\end{figure}

We find the $g$-factor from the lowest energy sublevel, because the statistical errors are lower for the lowest energy sublevel than for the upper one, as we have seen in Section \ref{sec-1}.
In Table   
  \ref{Table:g:rho} we collect the $g$ values obtained from the fit \eqref{eqLL1}, their errors, the lattice simulation parameters, the $\chi^2/n.d.f.$ values  and the field intervals used for the  fitting. In Fig. \ref{Fig:g:rho:mpi} all  $g$-factor values    are represented depending on the pion mass squared.
  
 \begin{table}[htb]
 \begin{center}
\begin{tabular}{c|r|r|r|r|r|r}
\hline
\hline
\rule{0cm}{0.4cm}
$V$     &   $m_{\pi}(\Mev)$&$a(\fm)$  &$g$-factor        &  $ \chi^2$/d.o.f.     & fit, $eB\, (\Gev^2)$ \\
\hline
 $18^4$    &   $331\pm 7$   & $0.115$  & $2.01\pm 0.18$   & $0.826$               & $[0,0.35]$  \\
 \hline
 $18^4$    &  $395\pm 6$   & $0.115$  & $2.17\pm 0.18$   & $0.969$               & $[0,0.35]$  \\
\hline
 $18^4$    &    $541\pm 3$    & $0.115$  & $2.12\pm 0.07$   & $1.159$               & $[0,0.35]$  \\
\hline
 $18^4$    &   $667\pm 3$   & $0.115$  & $2.07\pm 0.19$   & $1.695$               & $[0,0.35]$ \\
\hline
 $18^4$      &  $625\pm 21$   & $0.086$  &  $2.11\pm 0.01$  & $0.153$                & $[0,0.70]$ \\
 \hline
 $18^4$    &   $596\pm 12$   & $0.095$  & $2.30\pm 0.12$   & $1.094$                & $[0,0.55]$ \\
  \hline
 $18^4$     &  $572\pm 16$   & $0.105$  & $2.05\pm 0.03$   & $0.644$                & $[0,0.45]$\\
  \hline
 $20^4$     &  $535\pm 4$    & $0.115$  & $2.22\pm 0.08$   & $1.398$                & $[0,0.45]$\\
 \hline
 \hline
\end{tabular}
\end{center}
\caption{The g-factor of $\rho^{\pm}$ meson  for the various lattice volumes,   the pion masses and the   lattice spacings, the nonrenormalized quark masses are also shown. The  values of $\chi^2$/d.o.f. and fitting intervals are also represented.}
\label{Table:g:rho}
  \end{table}
 
 We consider several lattice spacings  to check the cut-off effects, is there any   $g$ value  dependence on the lattice spacing. 
The cut-off effects, as well as energy dependence on the pion mass, are   sizeable, as clearly
visible in the Fig.\ref{Fig:g:rho}, where results from different simulations do not lie on top of each other.  
Nevertheless, we did not observe any   dependence of the $g$-factor
on the lattice spacing or the pion mass within the error range, the relative systematic error  is $\sim 5\%$ and a mean value is equal to $2.1 \pm 0.1_{syst}$.
The finite volume effects are not so   pronounced, for example  lattices with     the volumes $18^4$ and $20^4$, the  lattice spacing $0.115\ \fm$ give very close  energy dependencies for the pion masses $541(3)$ and $535(4)$   respectively.
  
We have  obtained the most accurate value  $g=2.11\pm 0.01$ for the finest lattice with the spacing $a=0.086\ \fm$, the lattice volume $18^4$ and    the   pion mass $625(21)\ \Mev$.

 Our value of the $\rho^{\pm}$ magnetic moment  agrees with the other lattice predictions at   nonzero pion masses    \cite{Owen,Zanotti,Lee}. 
The magnetic moment in pure gauge theory differs from the full 2+1 QCD result,  its value is lower for approximately $ 5-10 \%$  for the same   pion masses \cite{Owen}.  This discrepancy may be explained by the contribution  of the dynamical quarks, cut-off and finite volume effects. 
 
D.G Gudino and G.T. Sanchez have obtained the g-factor from the analysis of BaBar cross section data for the reaction $e^{+}e^{-}\rightarrow \pi^+ \pi^- 2\pi^0$. They have  found the value  $g_{exp}=2.1\pm 0.5$  \cite{Gudino}.
The $g$-factor of the $\rho$ meson has been
calculated in \cite{Djukanovic} using the chiral EFT at  varying quark masses with a result similar to ours.
The light cone QCD sum rules predict the value  $g=2.4\pm 0.4$ \cite{Aliev}, and the covariant quark model gives   $g=2.14$ \cite{Melo}.

  \begin{figure}[htb]
\begin{center}
\includegraphics[width=5.8cm,angle=-90]{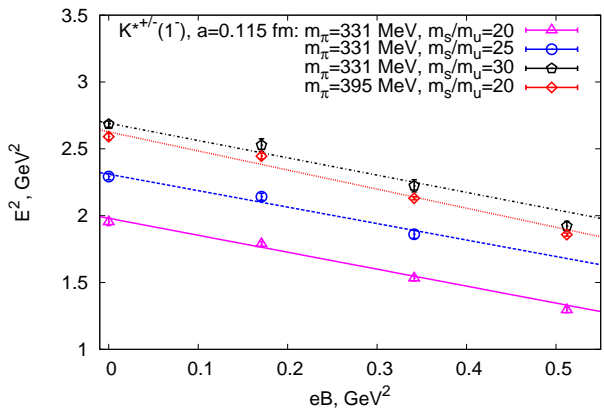}
\caption{The energy squared of $K^{*\pm}$ meson with the  spin projection 
$s_z=\pm1$ versus the magnetic field value for  the  lattice volume $18^4$, the lattice spacing $0.115\ \fm$ and various   light and  strange quark masses. Points correspond to the lattice data, lines are the fits to our data obtained using  function \eqref{eqLL1}.}
\label{Fig:g:K}
\end{center}
\end{figure}

 \begin{table}[htb]
 \begin{center}
\begin{tabular}{c|r|r|r|r}
\hline
\hline
\rule{0cm}{0.4cm}
  $m_{\pi}(\Mev)$& $m_s/m_u$   &$g$-factor  & $\chi^2$/d.o.f.  & fit, $eB\, (\Gev^2)$\\
\hline
  $331\pm 7$ & $20$     & $2.27 \pm 0.18$  & $1.845$          & $[0,0.35]$\\
\hline
  $331\pm 7$ & $25$     & $2.23 \pm 0.23$  & $1.986$          & $[0,0.35]$\\
\hline
  $331\pm 7$ & $30$     & $2.29 \pm 0.19$  & $1.366$          & $[0,0.35]$\\
\hline
\hline
\end{tabular}
\end{center}
\caption{The g-factor of the $K^{*\pm}$ meson, obtained on the  lattice with the volume $18^4$, the lattice spacing $0.115\ \fm$ and various bare   masses of the light and  strange quarks. The values of $\chi^2$/d.o.f. and intervals choice for the fitting procedure are also shown.}
\label{Table:g:K}
  \end{table}

  Fig.\ref{Fig:g:K} shows the energy squared of $K^{*\pm}$ meson with the spin projection $s_z=\pm 1$  for  the lightest pion mass $m_{\pi}=331(7)\ \Mev$, for the  strange quark mass  defined by the ratio $m_s/m_u=20,25,30$ and for pion mass  $m_{\pi}=395(6)\ \Mev,\ m_s=20m_u$. The fits to the lattice data, represented by the  lines, were obtained by using  formula \eqref{eqLL}. In Table \ref{Table:g:K}   the values of the g-factor for   various light and strange quark masses are collected.

 For the   pion mass $m_{\pi}=331(7)\ \Mev$ and  $m_s=30\, m_u$,  the closest to the physical case, the $g$-factor of the vector $K^{*\pm}$ meson is  $2.29 \pm 0.19$. This value is in accordance with the prediction of the QCD sum rules  $2.0\pm 0.4$ \cite{Aliev} and the  lattice results \cite{Zanotti,Lee}.
Any dependence of the g-factor on the strange quark mass has not been observed within the  error range.

The gyromagnetic ratio for free particles is close to "gravigyromagnetic" ratio, describing the motion of spin in rotating frames and near the rotating bodies  (see \cite{Teryaev:2016edw} and Ref. therein) which is universal ($g_G=2$) due to equivalence principle. The value of $g \approx 2$ for $\rho-$meson shows that QCD effects do not strongly violate this similarity. One may also refer to AdS QCD 
where g is exactly equal to 2 \cite{Grigoryan:2007vg}.

\section{Energy of $\rho^{\pm}$ meson}
\label{sec-3}

We have explored the ground state energy levels of the charged $\rho^{\pm}$ meson versus the value of the external magnetic field, which is directed along the 'z' axis.  

\begin{figure}[htb]
\begin{center}
\includegraphics[width=5.8cm,angle=-90]{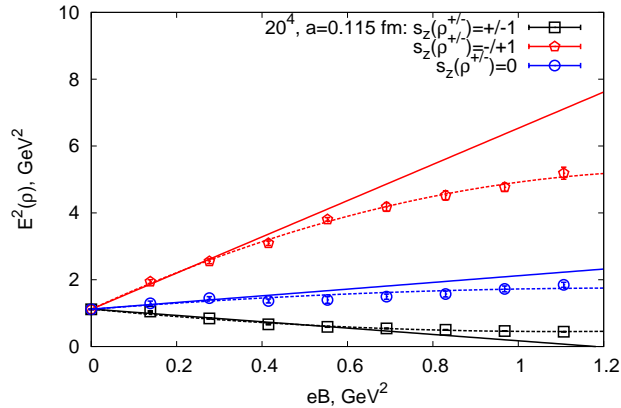}
\caption{The energy squared of the charged $\rho$ meson for  various spin projections $s_z=0,\pm 1$  depending on the magnetic field value for the lattice volume  $V=20^4$, the lattice spacing $a=0.115\ \fm$ and the     pion mass $m_{\pi^0}=535(4)\ \Mev$.
The solid lines correspond to the  fits of  the lattice data obtained with the use of   formula   \eqref{eqLL1}, the dashed lines are the fits of the lattice data by \eqref{eq:rho:s-1:B2}.}
\label{Fig:rho:LL}
\end{center}
\end{figure}

Fig.\ref{Fig:rho:LL}  shows the energy of the charged $\rho$ meson with
the  spin projections $s_z=-1,0$ and $+1$ on the  direction of the magnetic field depending on the  field value for
the lattice volume $20^4$, the lattice spacing $0.115\ \fm$ and the pion mass $m_{\pi}(B=0)=535(4)\ \Mev$. At the magnetic fields $eB\in[0,0.3]\ \Gev^2$ we fit the lattice data  by   formula \eqref{eqLL1}.  In Fig.\ref{Fig:rho:LL} the corresponding fits are   represented for the spin projections $s_z=-1,0$ and $+1$ by solid lines. 
  In our calculations the reversal of the field direction is equivalent to the replacement of a quark by an antiquark, so the  energy of the $\rho^-(s_z=-1)$ is identical to the energy of the $\rho^+(s_z=+1)$ meson.

At large magnetic fields the deviation from the Landau levels energy has been observed,
because the energy sublevels of the $\rho$ meson are likely getting sensitive on the internal structure of the meson,
which by definition is neglected when treating it as a free particle.

The parity has to be conserved, so for the spin projection $s_z=0$ on the field direction the  $\rho$-meson energy squared   contains only the terms of even powers of the magnetic field, while for the spin projections $1$ and $-1$ the terms of both odd and even powers of the  field are allowed.

In Fig.\ref{Fig:rho:LL}   we also show the fits of the lattice data (the dashed lines) by the following dependence
 \begin{equation}
E^2 =|qB| -gs_z qB +m^2 -4 \pi m \beta_m (qB)^2 
\label{eq:rho:s-1:B2}
\end{equation}
at  $eB\in[0,1.2]\ \Gev^2$, where $\beta_m$ is the dipole magnetic polarizability.

One can include the next power of the field into consideration and fit the data for the $s_z=\pm 1$ by the formula
\begin{equation}
E^2 =|qB| -gs_z qB +m^2 -4 \pi m \beta_m (qB)^2 - 4 \pi m \beta_m^{h1} (qB)^3,
\label{eq:rho:s-1:B3}
 \end{equation}
 where $\beta_m^{h1}$ is the magnetic hyperpolarizabilty of the first order.
We have found that the fit coincides with the one obtained with the use of formula \eqref{eq:rho:s-1:B2},  the contribution of the cubic term in the field is negligible in comparison with the quadratic term at $eB\in[0,1.2]$. The term of the fourth power in the magnetic field gives even smaller contributions to the energy of the vector meson for the all spin projections   at the field range considered, in this case the energy dependency   is the following
\begin{equation}
E^2 =|qB| -gs_z qB +m^2 -4 \pi m \beta_m (qB)^2 - 4 \pi m \beta_m^{h1} (qB)^3- 4 \pi m \beta_m^{h2} (qB)^4,
\label{eq:rho:s-1:B4}
 \end{equation}
 where $\beta_m^{h2}$  is the magnetic hyperpolarizabilty of the second order.
Thus, for the magnetic fields less than $1\ \Gev^2$ the energy squared  can be  expanded in a series in  the magnetic field. 
But at some value of the magnetic field   the series of the perturbation theory in the magnetic field begins to diverge and   the description in terms of the magnetic polarizabilities becomes poorly  defined. 
One of the main objectives pursued in this work is to find a  suitable range of field where the higher order effects can be neglected.

\section{Magnetic polarizabilities of $\rho$ meson}
\label{sec-4}

The magnetic polarizabilities in QCD, whose example is represented by (\ref{eq:rho:s-1:B3}),  quantify   the   response of an hadron  to the strong magnetic field. In case of the lightest $\pi$ meson this quantity can be found from the ChPT. But if we deal with the magnetic polarizability of $\rho$ meson, any theoretical and experimental predictions are difficult to obtain.  Thus  the lattice QCD  is a budding approach in this topic.

\begin{figure}[htb]
\begin{center}
\includegraphics[width=5.8cm,angle=-90]{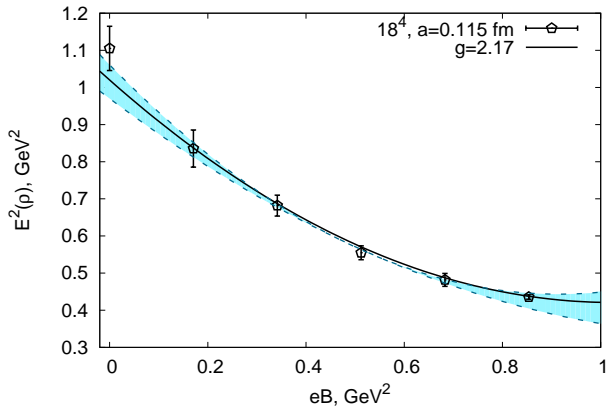}
\caption{The  energy squared of the ground state   of the $\rho^{+}$ ($\rho^{-}$) meson  with the  spin projection $s_z= + 1$ ($s_z= - 1$) versus the  field value for  the lattice spacing $a=0.115\ \fm$, the lattice volume $18^4$ and the pion mass  $395(6)\ \Mev$.  The curve is  the fit to the   lattice data performed of formula \eqref{eq:rho:s-1:B2}. The shaded area represents the error of $E^2$ at the fixed $g$-factor value.}
\label{Fig:rho:gfix1}
\end{center}
\end{figure}

We have found the magnetic dipole polarizabilities fitting the lattice data by  formula \eqref{eq:rho:s-1:B2}, where we fix the $g$-factor value, so that $m$ and $\beta_m$ are the  fit parameters, $qs_z=+1$.  The $g$-factor was found in Section \ref{sec-2} and presented in Table \ref{Table:g:rho}.  In Fig.\ref{Fig:rho:gfix1} the  solid line corresponds to the fit of the lattice data for the energy squared  at the average value of  the $g$-factor  for the lattice volume $18^4$, the lattice spacing $a=0.115\ \fm$ and the pion mass $395(6)\ \Mev$. The lattice data are shown by the empty points.  
The shaded area indicates the interval of the possible    energy values, if we change  the g-factor within the  error range. The  dashed  lines  correspond to the fits of the lattice data at $g=\bar{g}\pm \delta$, where $\delta$ is the error.

\begin{table}[htb]
 \begin{center}
\begin{tabular}{c|r|r|r|r|r}
\hline
\hline
\rule{0cm}{0.4cm}
$V$           & $m_{\pi}\,(\Mev)$ &$a\,(\fm)$    &$\beta_m\,(\Gev^{-3})$       & $ \chi^2/d.o.f$ & fit, $eB\, (\Gev^2)$\\
\hline
\rule{0cm}{0.4cm}
$18^4$           & $596 \pm 12$  & $0.095$    &   $-0.025^{+0.016}_{-0.014} $    & $1.656$      & $[0,0.9]$   \\
\hline
\rule{0cm}{0.4cm}
$18^4$           & $596 \pm 12$  & $0.095$    &   $-0.036^{+0.007}_{-0.006} $    & $1.864$      & $[0,1.3]$\\
\hline
\rule{0cm}{0.4cm}
 $18^4$          & $541 \pm 3$  & $0.115$    &   $-0.037^{+0.006}_{-0.005} $    & $2.774$        & $[0,1.05]$\\
\hline
\rule{0cm}{0.4cm}
 $20^4$          & $535 \pm 4$  & $0.115$    &   $-0.042^{+0.008}_{-0.008} $    & $2.274$        & $[0,1]$ \\
\hline
\rule{0cm}{0.4cm}
$18^4$           & $395 \pm 6$  & $0.115$    &   $-0.045^{+0.011}_{-0.012} $     &  $0.823$      & $[0,1]$ \\
\hline
\hline
\end{tabular}
\end{center}
\caption{The magnetic dipole polarizabilty $\beta_m$ of the charged $\rho$ meson for    the lattice spacings $0.095\ \fm$, $0.115\ \fm$,  for the lattice volume  $18^4$, various pion masses  and  for the lattice spacing $0.115\ \fm$, the lattice volume and the pion mass $535(4)\ \Mev$ with the corresponding errors and $\chi^2/d.o.f$. The fit interval  is  shown in the last column. The results were obtained from the 2-parametric fit according to formula \eqref{eq:rho:s-1:B2}.}
\label{Table_fixg_beta}
 \end{table}

The values of the magnetic dipole polarizability are represented in Table \ref{Table_fixg_beta}, the errors include both the deviation of the $\beta_m$  from its average value due to the uncertainty of the $g$-factor and the error of the fitting procedure.  For the lattice spacing $a=0.095\ \fm$  we show two  intervals of the magnetic field  which have been used for the determination of  the  polarizability, the results   agree with each other within the errors.

\begin{figure}[htb]
\begin{center}
\includegraphics[width=5.8cm,angle=-90]{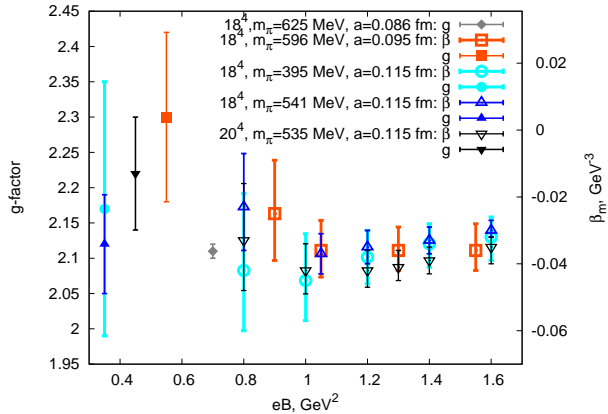}
\caption{The $g$-factor and the dipole magnetic polarizability $\beta_m$ depending on the interval of fields used for their determination for various lattices and pion masses.}
\label{Fig:beta:g:fit_int}
\end{center}
\end{figure}

 Fig. \ref{Fig:beta:g:fit_int} shows that the values of the $g$  -factor and the magnetic dipole polarizability $\beta_m$  slightly depend on the interval used for the fit procedure. With  the increase of the interval the $\beta_m$   value  starts to increase slowly from $eB\gtrsim 1 \Gev^2$. We explain this effect by the contribution of the magnetic hyperpolarizability  $\beta_m^{1h}$  \eqref{eq:rho:s-1:B3}, which gives an opposite contribution to the meson energy and not taken into account in equation \eqref{eq:rho:s-1:B2}.

\begin{figure}[htb]
\begin{center}
\includegraphics[width=5.8cm,angle=-90]{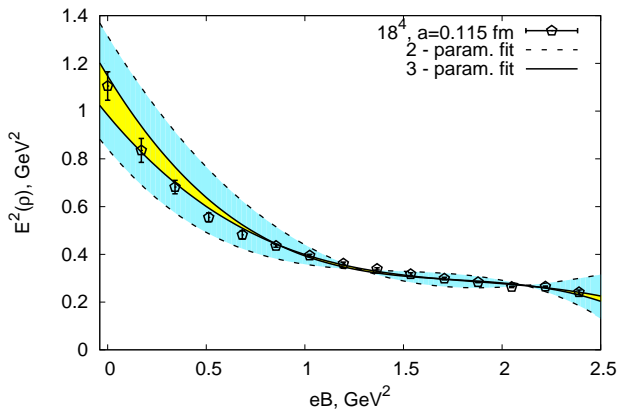}
\caption{The  energy squared of the ground state   of the $\rho^{\pm}$ mesons  with the  spin projection $s_z= \pm 1$  versus the  field value for  the lattice spacing $a=0.115\ \fm$, the lattice volume $18^4$ and the pion mass $395(6)\ \Mev$.  The solid curves restrict the area obtained using the 3-parametric fit \eqref{eq:rho:s-1:B3} at the fixed value of the $g$-factor. The dashed lines are the boundary of the acceptable $E^2$ values  if the $g$-factor and $\beta_m$ have the definite values.}
\label{Fig:rho:gfix2}
\end{center}
\end{figure}

We fit the energy squared using equation \eqref{eq:rho:s-1:B3} 
to obtain the magnetic hyperpolarizability of the first order  $\beta_m^{h1}$.
We can fix only  the $g$-factor or   the $g$-factor and the magnetic dipole polarizability $\beta_m$ at the same time.
 In the first case we  deal with the 3-parametric fit, so that $m$, $\beta_m$ and $\beta^{h1}_{m}$ are the fit parameters. In  Fig.\ref{Fig:rho:gfix2} the region of the acceptable $E^2$ values is restricted by   the solid lines, the corresponding values of the  $\beta_m$ and $\beta^{1h}_m$   are shown in Table \ref{3param_fit}.
  \begin{table}[htb]
 \begin{center}
\begin{tabular}{c|r|r|r|r|r|r}
\hline
\hline
\rule{0cm}{0.4cm}
$V$           & $m_{\pi}\,(\Mev)$ &$a\,(\fm)$    &$\beta_m\,(\Gev^{-3})$    &  $\beta_m^{1h}\,(\Gev^{-5})$  & $ \chi^2/d.o.f$ & $eB,\ \Gev^2$  \\
\hline
\rule{0cm}{0.4cm}
$18^4$           & $596 \pm 12$  & $0.095 $    &   $-0.050^{+0.009}_{-0.008} $ &  $0.009^{+0.002}_{-0.003}$   & $1.965$  &  $[0,2.5]$     \\
 \hline
 \rule{0cm}{0.4cm}
 $18^4$          & $541 \pm 3$  & $0.115$     &   $-0.045^{+0.005}_{-0.005} $ &  $0.009^{+0.001}_{-0.001}$    & $2.787$  &  $[0,2.5]$      \\
\hline
\rule{0cm}{0.4cm}
 $20^4$          & $535 \pm 4$  & $0.115$     &   $-0.058^{+0.008}_{-0.008} $  &  $0.013^{+0.002}_{-0.003}$ & $2.697$   &  $[0,2.0]$      \\
\hline
\rule{0cm}{0.4cm}
$18^4$           & $395 \pm 6$  & $0.115$     &   $-0.047^{+0.009}_{-0.009}$  &  $0.009^{+0.002}_{-0.002}$   &  $2.255$  &  $[0,2.5]$     \\
\hline
\hline
\end{tabular}
\end{center}
\caption{The magnetic dipole polarizabilty $\beta_m$, the magnetic hyperpolarizability $\beta^{1h}_m$ of the charged $\rho$ meson obtained from the 3-parametric fit \eqref{eq:rho:s-1:B3}, where we use the data presented in Table  \ref{Table:g:rho}. The results are obtained for various lattice volumes $V$, lattice spacings $a$ and pion masses $m_{\pi}$, $\chi^2/d.o.f.$ values correspond to the fit at the average $g$-factor value.}
\label{3param_fit}
 \end{table}

  \begin{table}[htb]
 \begin{center}
\begin{tabular}{c|r|r|r|r}
\hline
\hline
\rule{0cm}{0.4cm}
$V$           & $m_{\pi}\,(\Mev)$ &$a\,(\fm)$    &  $\beta_m^{1h}\,(\Gev^{-5})$  & $eB,\ \Gev^2$ \\
\hline
\rule{0cm}{0.4cm}
$18^4$           & $596 \pm 12$  & $0.095 $      &  $0.004^{+0.002}_{-0.002}$     &  $[0,2.5]$   \\
 \hline
 \rule{0cm}{0.4cm}
 $18^4$          & $541 \pm 3$  & $0.115$       &  $0.006^{+0.001}_{-0.002}$     &   $[0,2.5]$    \\
\hline
\rule{0cm}{0.4cm}
 $20^4$          & $535 \pm 4$  & $0.115$       &  $0.006^{+0.003}_{-0.002}$     &   $[0,2.0]$   \\
\hline
\rule{0cm}{0.4cm}
$18^4$           & $395 \pm 6$  & $0.115$      &  $0.008^{+0.005}_{-0.004}$      &   $[0,2.5]$  \\
\hline
\hline
\end{tabular}
\end{center}
\caption{The magnetic   hyperpolarizability $\beta^{1h}_m$ of the charged $\rho$ meson obtained from the 2-parametric fit \eqref{eq:rho:s-1:B3}, where we use the data from   Table \ref{Table:g:rho} and  Table \ref{Fig:g:rho}. The results are obtained for   various lattice volumes $V$,   lattice spacings $a$ and  pion  masses $m_{\pi}$.}
\label{2param_fit}
 \end{table}
  Zero hyperpolarizability of the second order  $\beta_m^{h2}$  was obtained   within the errors fitting  the  data   according to formula \eqref{eq:rho:s-1:B4}  
  at $eB\in [0,2.5]\ \Gev^2$, so we consider its contribution  insignificant.

  We also  have found the magnetic hyperpolarizability $\beta^{h1}_{m}$ from the 2-parametric fit \eqref{eq:rho:s-1:B3}   for a check, using the fixed  $g$-factor from Table \ref{Table:g:rho} and the      $\beta_m$ from   Table    \ref{Table_fixg_beta}. In Fig. \ref{Fig:rho:gfix2} the region limited by the dashed lines corresponds to such fits  if we change the $g$-factor and the magnetic dipole polarizability  $\beta_m$   within their errors. The values of  $\beta^{h1}_{m}$  are presented in Table \ref{2param_fit} and agree with the results obtained with the use of the 3-parametric fit.

At the value of the magnetic dipole polarizability $\sim 0.045\ \Gev^{-3}$   the contribution  of the quadratic term in  field to the energy squared  is approximately equal to $10\%$ at  $eB= 0.35\ \Gev^2$, that is at least not larger  than the errors of the $g$-factor.   
At the magnetic field $\sim 1\ \Gev^2$ the term with the hyperpolarizability gives the correction $\sim 20\%$ to the term with   the magnetic dipole polarizability, which is compatible with its errors. The term with hyperpolarizability  partially compensates  the quadratic term in field   at the magnetic fields considered here.

\begin{figure}[htb]
\begin{center}
\includegraphics[width=5.8cm,angle=-90]{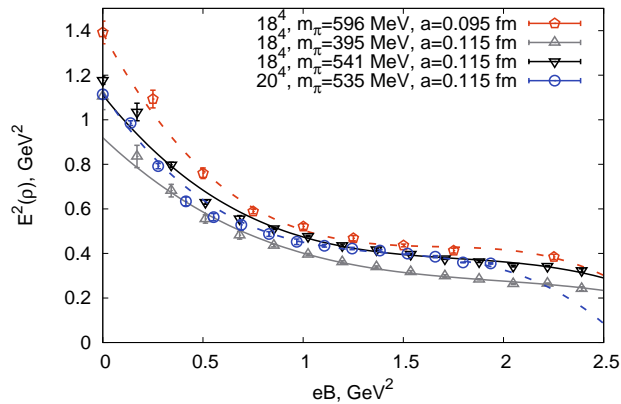}
\caption{The  energy squared of the ground state   of the $\rho^{+}$ ($\rho^{-}$)  meson  with the  spin projection $s_z= +1$ ($s_z= -1$) versus the  field value for  various sets of lattice data. The curves are  the fits to the  lattice data obtained with the use of  formula \eqref{eq:rho:s-1:B3}.}
\label{Fig:rho:s-1}
\end{center}
\end{figure}

In Fig.\ref{Fig:rho:s-1} the  energy squared of the ground state of the vector $\rho^{\pm}$ meson with the
spin projection  $s_z=\pm1$  is shown  by points for the various lattice data sets. The solid and dashed lines are the 4-parametric fits to the lattice data obtained with the use of formula \eqref{eq:rho:s-1:B3}, where $m$, $g$, $\beta_m$ and $\beta^{1h}_m$ are the fit parameters presented in Table \ref{Table4param}.
These results agree with the previous results for the $g$-factor, magnetic polarizability and hyperpolarizability, but the $g$-factor values have bigger errors.

 \begin{table}[htb]
 \begin{center}
\begin{tabular}{c|r|r|r|r|r|r}
\hline
\hline
$V$              & $m_{\pi}(\Mev)$    &$a(\fm)$ &  $g$-factor    & $\beta_m(\Gev^{-3})$  & $\beta^{1h}_m(\Gev^{-5})$ & $\chi^2/d.o.f.$\\
\hline
$18^4$           & $596\pm 12$  &       $0.095$ &    $2.49 \pm 0.18$  & $-0.056\pm 0.008$     &  $0.010 \pm 0.002 $  & $3.738$ \\
\hline
 $18^4$          & $541 \pm 3$  &       $0.115$ &   $2.09\pm 0.09$   &  $-0.043\pm 0.005$     &  $0.008\pm 0.001$  & $2.993$ \\
\hline
 $20^4$          & $535 \pm 4$  &      $0.115$  &    $2.35\pm 0.11$     &  $-0.067\pm0.008$     & $0.015\pm0.002$   & $2.657$ \\
\hline
$18^4$           & $395 \pm 6$  &      $0.115$  &    $1.85\pm0.11 $   &  $ -0.033\pm 0.006$    & $0.006\pm 0.001$   & $3.191$  \\
\hline
\hline
\end{tabular}
\end{center}
\caption{The magnetic dipole moment, the magnetic dipole polarizability and magnetic hyperpolarizability of the first   order $\beta^{1h}_m$ and second order $\beta^{2h}_m$  of the charged $\rho$ meson for  the lattice spacings $0.095\ \fm$, $0.115\ \fm$, the lattice volume    $18^4$, various pion masses and for the lattice spacing $0.115\ \fm$, the lattice volume $20^4$ and the pion mass $m_{\pi}=535(4)\  \Mev$  with their errors  and $ \chi^2/d.o.f$ values. The results were obtained with the use of 4-parametric fit \eqref{eq:rho:s-1:B3}.}
\label{Table4param}
 \end{table}

The nonlinear terms containing  magnetic polarizabilities make  the energy dependence rather flat at large magnetic fields, otherwise the lowest energy sublevel would cross zero at $eB\sim 1\ \Gev^2$ as it follows from  the Landau level picture  \eqref{eqLL1}.
The higher order dependence of the energy on $B$ tends to be sign-changing from order to order and   results in rather flat behaviour.
This is quite important and constitutes one of the main results of the present paper. 

The importance stems from the fact that 
 the lowest order linear dependence leads to the node of the mass square of the $\rho^- (\rho^+)$ meson for $s_=-1(s_z=1)$.
This signals the appearance of tachyonic mode which may lead to superconducting phase transition and Abrikosov lattice 
appearance  known for many years in electroweak theory \cite{Skalozub:1986be}. The transition to superconducting phase was recently discovered and extensively studied in  the actual case of $\rho$ mesons (See \cite{Chernodub} and Ref. therein). 

We observe that the magnetic (hyper)polarizabilities provide the "self-healing" 
of such a behaviour. One may even speculate, that QCD prevents the phase transition and saves our Universe in the case of presence of even extremely large magnetic fields at early stage of its evolution.  At the same time, one cannot exclude, that the phase transition may emerge in some other manner.

\begin{figure}[htb]
\begin{center}
\includegraphics[width=5.8cm,angle=-90]{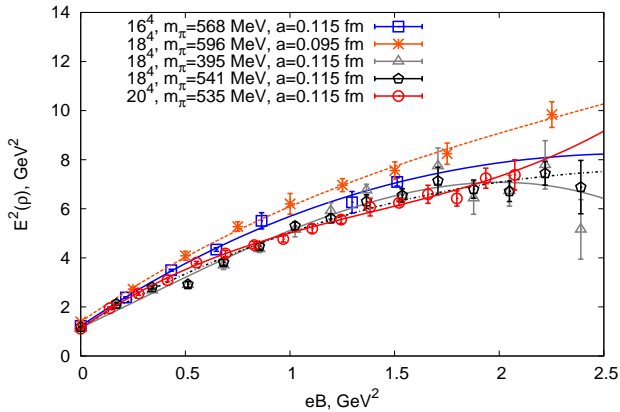}
\caption{The  energy squared of the ground state   of $\rho^-$ ($\rho^+$) meson  with   $s_z= +1$ ($s_z=-1$)  depending on the magnetic  field value for   lattice volumes $16^4$, $18^4$, $20^4$,  lattice spacings $0.095\ \fm$, $0.115\ \fm$ and  various pion masses  with the fits made using  formula \eqref{eq:rho:s-1:B3}.}
\label{Fig:rho:s+1}
\end{center}
\end{figure}

Fig.\ref{Fig:rho:s+1} shows the energy squared of vector $\rho^-$ ($\rho^+$)  meson with the spin projection $s_z=+1$ ($s_z=-1$)  as a function of the  field value. The $E^2$ behaviour is discribed by formula \eqref{eq:rho:s-1:B3} at $qs_z=-1$. According to the symmetries the absolute values of the hyperpolarizabilities are equal for the $s_z=-1$ and $s_z=+1$ cases, so the term  of the fourth power in $B$ has to give a very small relative contribution to the energy at  the considered range of fields. 

The extrapolation errors also appear to be essentially larger for the highest energy sublevel with the spin orientation opposite to magnetic field direction resulting in the higher energy. As a result, the values of the magnetic moment and the (hyper)polarizability are compatible within errors for two energy sublevels, while the lower energy sublevel providing the smaller errors is used for obtaining final results for these physical quantities. One can see from Fig.\ref{Fig:rho:s+1} that there are no finite volume effects for this energy component, and energy data lie higher for $16^4$ lattice volume  than for $18^4$ and $20^4$ due to a bit larger pion mass.

\begin{figure}[htb]
\begin{center}
\includegraphics[width=5.8cm,angle=-90]{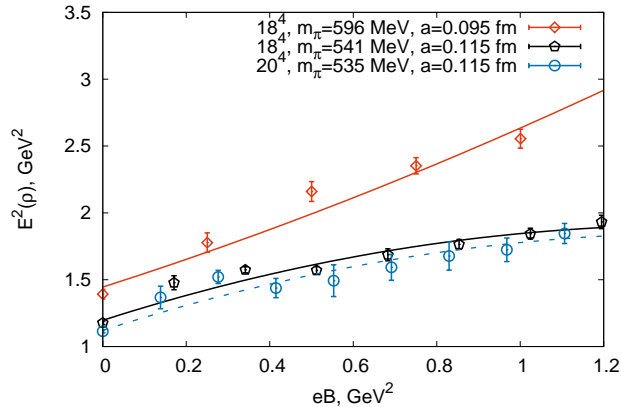}
\caption{The  energy squared of the ground state   of the  $\rho^{\pm}$ meson  with  the spin projection $s_z= 0$ versus the field value for  the lattice volumes  $18^4$ and $20^4$, the lattice spacings $0.095\ \fm$, $0.115\ \fm$ and various pion masses $m_{\pi}$  with the corresponding fits   \eqref{eq:rho:s0}.}
\label{Fig:rho:s0}
\end{center}
\end{figure}

In Fig.\ref{Fig:rho:s0}  we show the energy squared of the  charged $\rho$ meson with the $s_z=0$   versus the field value for various lattices. Due to the parity conservation   the nonlinear terms in $B$ can give contributions to the $E^2$ value    if they contain only  even powers of the  field.
 So, we fit our lattice data by the following formula  
 \begin{equation}
E^2 =|qB|  +m^2 -4 \pi m \beta_m (qB)^2,
\label{eq:rho:s0}
\end{equation}
at $eB\in[0,1.2]\Gev^2$, where $m$, $\beta_m$   are the fit parameters.
Statistically significant values of polarizabilities were found for the lattices with the  volumes $18^4$ and $20^4$, the lattice spacing $0.115\ \fm$ and the quark mass $34.26\ \Mev$,   the results are shown in Table \ref{Table_s0}.
For   all the  sublevels  the energy of the $\rho$ meson is   dependent on the lattice spacing and the pion mass.   

  \begin{table}[htb]
 \begin{center}
\begin{tabular}{c|r|r|r|r}
\hline
\hline
$V$  & $m_{\pi}(\Mev)$ &$a(\fm)$           & $\beta_m(\Gev^{-3})$    &   $\chi^2/d.o.f.$ \\
\hline
$18^4$           & $541 \pm 3$  & $0.115$   & $0.026\pm 0.004$        &  $1.959$ \\
\hline
 $20^4$          & $535 \pm 4$  & $0.115$   & $0.026\pm 0.005$           &   $1.365$ \\
\hline
\hline
\end{tabular}
\end{center}
\caption{The magnetic dipole polarizability  of the charged $\rho$ meson obtained for the      lattice spacing $0.115\ \fm$ and the lattice volumes 
  $18^4$, $20^4$ from the fit \ref{eq:rho:s0}.  The pion masses  are represented in the second column, the values of $ \chi^2/d.o.f$   are shown in the last one.}
 \label{Table_s0}
 \end{table}

  Our results qualitatively disagree from the findings of \cite{Hidaka}, where at small value of B the energy of the lowest energy sublevel is observed to first decrease, then increase again.     The possible explanation is that the overlap fermions do not break the full chiral symmetry in contrast to  
Wilson fermions. It enable to avoid the additional term for  the renormalised quark mass   which is not zero for Wilson fermions and depends on the lattice spacing and  the magnetic field value  \cite{Bali:2015,Bali2}. 
Besides, the  extraction procedure of the correlation functions for different spin  projections is described in  \cite{Hidaka} briefly, so the detailed comparison is currently difficult to perform.

\section{Conclusion}

The energy levels of the ground state of vector $\rho^{\pm}$ and $K^{*\pm}$ mesons have been explored versus the magnetic field value in the $SU(3)$ quenched lattice gauge theory. At low magnetic fields our data agrees with the picture of the  Landau levels within the error range. At large magnetic fields  $eB>[0.3\div 0.5]\, \Gev^2$ the non-linear terms of the magnetic field give  a contribution  to the energy providing  clear indications of the nonzero magnetic dipole polarizability and the hyperpolarizabilities. Our lattice data for the energy squared confirms the theoretical expectations.

We have calculated the magnetic dipole polarizability and hyperpolarizability of the charged  $\rho$ mesons, while for the $s_z=0$ only the dipole polarizability has been found.  For the low energy component we do not observe any significant dependence of the results on the lattice spacing and the lattice volume. But we expect these  effects   can appear for the highest energy level.     
 
We  have also found the g-factor of the vector    $\rho^{\pm}$ and $K^{*\pm}$ mesons. The obtained  values  agree with the   predictions of the QCD sum rules \cite{Aliev}, the covariant quark model  \cite{Melo} and the lattice results at close pion masses \cite{Zanotti,Lee}. The magnetic moment of the  $\rho^{\pm}$ meson is in good agreement with the experimental result \cite{Gudino}.

The magnetic dipole polarizability of the  $\rho^{\pm}$ meson for  the
$s_z=-1$ and $s_z=+1$ energy components   has to be the same, and we
found it from the lowest energy sublevel with a good accuracy. Also we
calculate the $\beta_m$ value for the $s_z=0$ case, it's opposite in sign and its absolute value is  lower than for the $|s_z|=1$ for the same lattice data sets.    The polarizability tensor
of the $\rho^{\pm}$ meson at nonzero momentum is of a particular interest  because it could be
associated with the lepton asymmetry in a strong magnetic field \cite{Buividovich:2012kq}.

It has  been  shown, that the meson energy doesn't turn to zero due to higher order dependence of the energy from B. The tachyonic mode does not seem to exist in QCD at zero temperature and chemical potential, and there is no respective condensation of $\rho$ mesons as it was predicted in some theoretical works, although one cannot exclude another mechanism of condensation.

\acknowledgments

We are thankful to Yu. A. Simonov and V.V. Skalozub for useful discussions. 
   The authors are grateful to  FAIR-ITEP supercomputer center where these numerical calculations were performed.
   This work   is completely supported by a grant from the Russian Science Foundation (project number 16-12-10059).
   
   \newpage

\appendix
\section{Lattice parameters and values of the magnetic field}
\label{app}
 \begin{table}[htb]
  \caption{The lattice parameters  such as the volume, the lattice spacing, the pion mass and the number of the lattice gauge configurations used for the calculations for the each value of the magnetic field. The value of the magnetic can be calculated according to equation (2.13), the negative value of $n_B$ corresponds to the negative projection of the magnetic field on the 'z' axis.}
 \begin{center}
\begin{tabular}{|c|r|r|r|r|}
\hline
\hline
\rule{0cm}{0.4cm}
$V$        & $a(\fm)$ & $m_{\pi}(\Mev)$   & $n_B$    &  $N_{conf}$  \\
\hline
 $18^4$    & $0.086$  &   $625\pm 21$     & $0$             & $199$  \\
 \hline
 $18^4$    & $0.086$  &   $625\pm 21$     & $1$            & $199$  \\
 \hline
 $18^4$    & $0.086$  &   $625\pm 21$     & $2$             & $199$  \\
 \hline
 $18^4$    & $0.086$  &   $625\pm 21$     & $-2$             & $198$  \\
 \hline
 $18^4$    & $0.086$  &   $625\pm 21$     & $-4$             & $199$  \\
 \hline
  $18^4$    & $0.095$  &   $596\pm 12$     & $0$            & $198$  \\
 \hline
  $18^4$    & $0.095$  &   $596\pm 12$     & $1$             & $294$  \\
   \hline
  $18^4$    & $0.095$  &   $596\pm 12$     & $2$             & $295$  \\
   \hline
  $18^4$    & $0.095$  &   $596\pm 12$     & $3$             & $199$  \\
   \hline
  $18^4$    & $0.095$  &   $596\pm 12$     & $4$             & $198$  \\
   \hline
  $18^4$    & $0.095$  &   $596\pm 12$     & $5$             & $194$  \\
   \hline
  $18^4$    & $0.095$  &   $596\pm 12$     & $6$             & $196$  \\
   \hline
  $18^4$    & $0.095$  &   $596\pm 12$     & $7$             & $197$  \\
   \hline
  $18^4$    & $0.095$  &   $596\pm 12$     & $8$             & $198$  \\
   \hline
  $18^4$    & $0.095$  &   $596\pm 12$     & $9$             & $198$  \\
   \hline
  $18^4$    & $0.095$  &   $596\pm 12$     & $10$             & $198$  \\
   \hline
  $18^4$    & $0.095$  &   $596\pm 12$     & $11$            & $198$  \\
   \hline
  $18^4$    & $0.095$  &   $596\pm 12$     & $12$             & $197$  \\
   \hline
  $18^4$    & $0.095$  &   $596\pm 12$     & $13$             & $196$  \\
   \hline
  $18^4$    & $0.095$  &   $596\pm 12$     & $-1$            & $295$  \\
   \hline
  $18^4$    & $0.095$  &   $596\pm 12$     & $-2$            & $295$  \\
   \hline
  $18^4$    & $0.095$  &   $596\pm 12$     & $-4$           & $199$  \\
   \hline
  $18^4$    & $0.095$  &   $596\pm 12$     & $-6$              & $199$  \\
   \hline
  $18^4$    & $0.095$  &   $596\pm 12$     & $-8$            & $198$  \\
   \hline
  $18^4$    & $0.095$  &   $596\pm 12$     & $-10$             & $199$  \\
   \hline
  $18^4$    & $0.095$  &   $596\pm 12$     & $-12$            & $197$  \\
   \hline
  $18^4$    & $0.095$  &   $596\pm 12$     & $-14$             & $198$  \\
   \hline
  $18^4$    & $0.095$  &   $596\pm 12$     & $-16$             & $198$  \\
   \hline
  $18^4$    & $0.095$  &   $596\pm 12$     & $-18$            & $197$  \\
   \hline
  $18^4$    & $0.095$  &   $596\pm 12$     & $-20$             & $198$  \\
   \hline
  $18^4$    & $0.095$  &   $596\pm 12$     & $-22$             & $197$  \\
   \hline
  $18^4$    & $0.095$  &   $596\pm 12$     & $-24$             & $199$  \\
    \hline
   \hline
\end{tabular}
\end{center}
 \end{table}

 \begin{center}
\begin{tabular}{|c|r|r|r|r|}
\hline
\hline
\rule{0cm}{0.4cm}
$V$        & $a(\fm)$ & $m_{\pi}(\Mev)$   & $n_B$           &  $N_{conf}$ \\
   \hline
  $18^4$    & $0.095$  &   $596\pm 12$     & $-26$             & $197$  \\
   \hline
  $18^4$    & $0.115$  &   $331\pm 7 $     & $0$               & $336$  \\
     \hline
  $18^4$    & $0.115$  &   $331\pm 7 $     & $1$               & $271$  \\
     \hline
  $18^4$    & $0.115$  &   $331\pm 7 $     & $2$              & $337$  \\
     \hline
  $18^4$    & $0.115$  &   $331\pm 7 $     & $3$             & $239$  \\
     \hline
  $18^4$    & $0.115$  &   $331\pm 7 $     & $-2$             & $277$  \\
   \hline
  $18^4$    & $0.115$  &   $331\pm 7 $     & $-4$             & $274$  \\
     \hline
  $18^4$    & $0.115$  &   $331\pm 7 $     & $-6$             & $238$  \\
        \hline
  $18^4$    & $0.115$  &   $395\pm 6 $     & $0$            & $245$  \\
    \hline
  $18^4$    & $0.115$  &   $395\pm 6 $     & $1$             & $235$  \\
    \hline
  $18^4$    & $0.115$  &   $395\pm 6 $     & $2$             & $245$  \\
    \hline
  $18^4$    & $0.115$  &   $395\pm 6 $     & $3$             & $332$  \\
    \hline
  $18^4$    & $0.115$  &   $395\pm 6 $     & $4$            & $246$  \\
     \hline
       $18^4$    & $0.115$  &   $395\pm 6 $     & $5$             & $250$  \\
    \hline
  $18^4$    & $0.115$  &   $395\pm 6 $     & $6$             & $249$  \\
    \hline
  $18^4$    & $0.115$  &   $395\pm 6 $     & $7$            & $250$  \\
    \hline
  $18^4$    & $0.115$  &   $395\pm 6 $     & $8$            & $336$  \\
    \hline
  $18^4$    & $0.115$  &   $395\pm 6 $     & $9$            & $336$  \\
    \hline
  $18^4$    & $0.115$  &   $395\pm 6 $     & $10$            & $335$  \\
    \hline
  $18^4$    & $0.115$  &   $395\pm 6 $     & $11$             & $248$  \\
    \hline
  $18^4$    & $0.115$  &   $395\pm 6 $     & $12$             & $244$  \\
    \hline
  $18^4$    & $0.115$  &   $395\pm 6 $     & $13$           & $249$  \\
    \hline
  $18^4$    & $0.115$  &   $395\pm 6 $     & $14$            & $250$  \\
    \hline
  $18^4$    & $0.115$  &   $395\pm 6 $     & $-2$           & $249$  \\
    \hline
  $18^4$    & $0.115$  &   $395\pm 6 $     & $-4$            & $250$  \\
    \hline
  $18^4$    & $0.115$  &   $395\pm 6 $     & $-6$           & $311$  \\
    \hline
  $18^4$    & $0.115$  &   $395\pm 6 $     & $-8$           & $198$  \\
    \hline
  $18^4$    & $0.115$  &   $395\pm 6 $     & $-10$            & $241$  \\
    \hline
  $18^4$    & $0.115$  &   $395\pm 6 $     & $-12$            & $232$  \\
    \hline
  $18^4$    & $0.115$  &   $395\pm 6 $     & $-14$           & $238$  \\
    \hline
  $18^4$    & $0.115$  &   $395\pm 6 $     & $-16$            & $320$  \\
    \hline
  $18^4$    & $0.115$  &   $395\pm 6 $     & $-18$            & $333$  \\
    \hline
  $18^4$    & $0.115$  &   $395\pm 6 $     & $-20$          & $336$  \\
    \hline
  $18^4$    & $0.115$  &   $395\pm 6 $     & $-22$            & $248$  \\
    \hline
  $18^4$    & $0.115$  &   $395\pm 6 $     & $-24$           & $247$  \\
    \hline
  $18^4$    & $0.115$  &   $395\pm 6 $     & $-26$            & $248$  \\
    \hline
  $18^4$    & $0.115$  &   $395\pm 6 $     & $-28$            & $245$  \\
  \hline
  $18^4$    & $0.115$  &   $541\pm 3 $     & $0$            & $294$  \\
    \hline
  $18^4$    & $0.115$  &   $541\pm 3 $     & $1$            & $335$  \\
    \hline
  $18^4$    & $0.115$  &   $541\pm 3 $     & $2$            & $335$  \\
    \hline
  $18^4$    & $0.115$  &   $541\pm 3 $     & $3$            & $249$  \\
    \hline
  $18^4$    & $0.115$  &   $541\pm 3 $     & $4$            & $304$  \\
   \hline
   \hline
\end{tabular}
\end{center}

 \begin{center}
\begin{tabular}{|c|r|r|r|r|}
\hline
\hline
\rule{0cm}{0.4cm}
$V$        & $a(\fm)$ & $m_{\pi}(\Mev)$   & $n_B$           &  $N_{conf}$ \\
    \hline
  $18^4$    & $0.115$  &   $541\pm 3 $     & $5$            & $298$  \\
    \hline
  $18^4$    & $0.115$  &   $541\pm 3 $     & $6$            & $248$  \\
    \hline
  $18^4$    & $0.115$  &   $541\pm 3 $     & $7$            & $298$  \\
     \hline
  $18^4$    & $0.115$  &   $541\pm 3 $     & $8$            & $299$  \\
    \hline
  $18^4$    & $0.115$  &   $541\pm 3 $     & $9$            & $300$  \\
    \hline
  $18^4$    & $0.115$  &   $541\pm 3 $     & $10$            & $299$  \\
     \hline
    $18^4$    & $0.115$  &   $541\pm 3 $     & $11$            & $248$  \\
   \hline
    $18^4$    & $0.115$  &   $541\pm 3 $     & $12$            & $250$  \\
      \hline
 $18^4$    & $0.115$  &   $541\pm 3 $     & $13$            & $249$  \\
   \hline
    $18^4$    & $0.115$  &   $541\pm 3 $     & $14$            & $250$  \\
   \hline
    $18^4$    & $0.115$  &   $541\pm 3 $     & $-1$            & $213$  \\
   \hline
    $18^4$    & $0.115$  &   $541\pm 3 $     & $-2$            & $250$  \\
   \hline
    $18^4$    & $0.115$  &   $541\pm 3 $     & $-3$            & $250$  \\
   \hline
    $18^4$    & $0.115$  &   $541\pm 3 $     & $-4$            & $335$  \\
   \hline
    $18^4$    & $0.115$  &   $541\pm 3 $     & $-5$            & $250$  \\
   \hline
    $18^4$    & $0.115$  &   $541\pm 3 $     & $-6$            & $249$  \\
   \hline
    $18^4$    & $0.115$  &   $541\pm 3 $     & $-7$            & $250$  \\
   \hline
    $18^4$    & $0.115$  &   $541\pm 3 $     & $-8$            & $309$  \\
   \hline
    $18^4$    & $0.115$  &   $541\pm 3 $     & $-9$            & $234$  \\
   \hline
    $18^4$    & $0.115$  &   $541\pm 3 $     & $-10$            & $297$  \\
   \hline
    $18^4$    & $0.115$  &   $541\pm 3 $     & $-11$            & $250$  \\
   \hline
    $18^4$    & $0.115$  &   $541\pm 3 $     & $-12$            & $249$  \\
   \hline
    $18^4$    & $0.115$  &   $541\pm 3 $     & $-13$            & $250$  \\
   \hline
    $18^4$    & $0.115$  &   $541\pm 3 $     & $-14$            & $298$  \\
   \hline
    $18^4$    & $0.115$  &   $541\pm 3 $     & $-16$            & $300$  \\
   \hline
    $18^4$    & $0.115$  &   $541\pm 3 $     & $-18$            & $300$  \\
   \hline
    $18^4$    & $0.115$  &   $541\pm 3 $     & $-20$            & $298$  \\
   \hline
    $18^4$    & $0.115$  &   $541\pm 3 $     & $-22$            & $249$  \\
   \hline
    $18^4$    & $0.115$  &   $541\pm 3 $     & $-24$            & $250$  \\
   \hline
    $18^4$    & $0.115$  &   $541\pm 3 $     & $-26$            & $250$  \\
   \hline
    $18^4$    & $0.115$  &   $541\pm 3 $     & $-28$            & $247$  \\
    \hline
    $18^4$    & $0.115$  &   $ 667 \pm 3  $     & $0$            & $290$  \\
     \hline
    $18^4$    & $0.115$  &   $ 667 \pm 3  $     & $1$            & $195$  \\
     \hline
    $18^4$    & $0.115$  &   $ 667 \pm 3  $     & $2$            & $196$  \\
     \hline
    $18^4$    & $0.115$  &   $ 667 \pm 3  $     & $3$            & $266$  \\
     \hline
    $18^4$    & $0.115$  &   $ 667 \pm 3  $     & $4$            & $199$  \\
     \hline
    $18^4$    & $0.115$  &   $ 667 \pm 3  $     & $-1$            & $198$  \\
     \hline
    $18^4$    & $0.115$  &   $ 667 \pm 3  $     & $-2$            & $199$  \\
     \hline
    $18^4$    & $0.115$  &   $ 667 \pm 3  $     & $-3$            & $300$  \\
     \hline
    $18^4$    & $0.115$  &   $ 667 \pm 3  $     & $-4$            & $197$  \\
       \hline
    $20^4$    & $0.115$  &   $ 535 \pm 4  $     & $0$            & $290$  \\
       \hline
    $20^4$    & $0.115$  &   $ 535 \pm 4  $     & $1$            & $299$  \\
       \hline
   \hline
\end{tabular}
\end{center}

 \begin{center}
\begin{tabular}{|c|r|r|r|r|}
\hline
\hline
\rule{0cm}{0.4cm}
$V$        & $a(\fm)$ & $m_{\pi}(\Mev)$   & $n_B$           &  $N_{conf}$ \\
        \hline
    $20^4$    & $0.115$  &   $ 535 \pm 4  $     & $2$            & $295$  \\
        \hline
    $20^4$    & $0.115$  &   $ 535 \pm 4  $     & $3$            & $291$  \\
         \hline
    $20^4$    & $0.115$  &   $ 535 \pm 4  $     & $4$            & $296$  \\
        \hline
    $20^4$    & $0.115$  &   $ 535 \pm 4  $     & $5$            & $198$  \\
        \hline
    $20^4$    & $0.115$  &   $ 535 \pm 4  $     & $6$            & $296$  \\
        \hline
    $20^4$    & $0.115$  &   $ 535 \pm 4  $     & $7$            & $289$  \\
        \hline
    $20^4$    & $0.115$  &   $ 535 \pm 4  $     & $8$            & $298$  \\
      \hline
      $20^4$    & $0.115$  &   $ 535 \pm 4  $     & $9$            & $287$  \\
        \hline
    $20^4$    & $0.115$  &   $ 535 \pm 4  $     & $10$            & $294$  \\
        \hline
    $20^4$    & $0.115$  &   $ 535 \pm 4  $     & $11$            & $100$  \\
        \hline
    $20^4$    & $0.115$  &   $ 535 \pm 4  $     & $12$            & $100$  \\
        \hline
    $20^4$    & $0.115$  &   $ 535 \pm 4  $     & $13$            & $100$  \\
        \hline
    $20^4$    & $0.115$  &   $ 535 \pm 4  $     & $14$            & $100$  \\
        \hline
    $20^4$    & $0.115$  &   $ 535 \pm 4  $     & $-2$            & $278$  \\
        \hline
    $20^4$    & $0.115$  &   $ 535 \pm 4  $     & $-4$            & $295$  \\
          \hline
      $20^4$    & $0.115$  &   $ 535 \pm 4  $     & $-6$            & $295$  \\
        \hline
    $20^4$    & $0.115$  &   $ 535 \pm 4  $     & $-8$            & $256$  \\
        \hline
    $20^4$    & $0.115$  &   $ 535 \pm 4  $     & $-10$            & $252$  \\
        \hline
    $20^4$    & $0.115$  &   $ 535 \pm 4  $     & $-12$            & $300$  \\
        \hline
    $20^4$    & $0.115$  &   $ 535 \pm 4  $     & $-14$            & $298$  \\
        \hline
    $20^4$    & $0.115$  &   $ 535 \pm 4  $     & $-16$            & $300$  \\
        \hline
    $20^4$    & $0.115$  &   $ 535 \pm 4  $     & $-18$            & $300$  \\
        \hline
    $20^4$    & $0.115$  &   $ 535 \pm 4  $     & $-20$            & $300$  \\
          \hline
      $20^4$    & $0.115$  &   $ 535 \pm 4  $     & $-22$            & $100$  \\
        \hline
    $20^4$    & $0.115$  &   $ 535 \pm 4  $     & $-24$            & $99$  \\
        \hline
    $20^4$    & $0.115$  &   $ 535 \pm 4  $     & $-26$            & $100$  \\
        \hline
    $20^4$    & $0.115$  &   $ 535 \pm 4  $     & $-28$            & $96$  \\
       \hline
   \hline
\end{tabular}
\end{center}

\end{document}